\documentclass{moriond}

\def\Journal#1#2#3#4{{#1} {\bf #2}, #3 (#4)}

\def\be{\begin{equation}}
\def\ee{\end{equation}}
\def\bea{\begin{eqnarray}}
\def\eea{\end{eqnarray}}

\newcommand{\Photo}{\includegraphics[height=35mm]{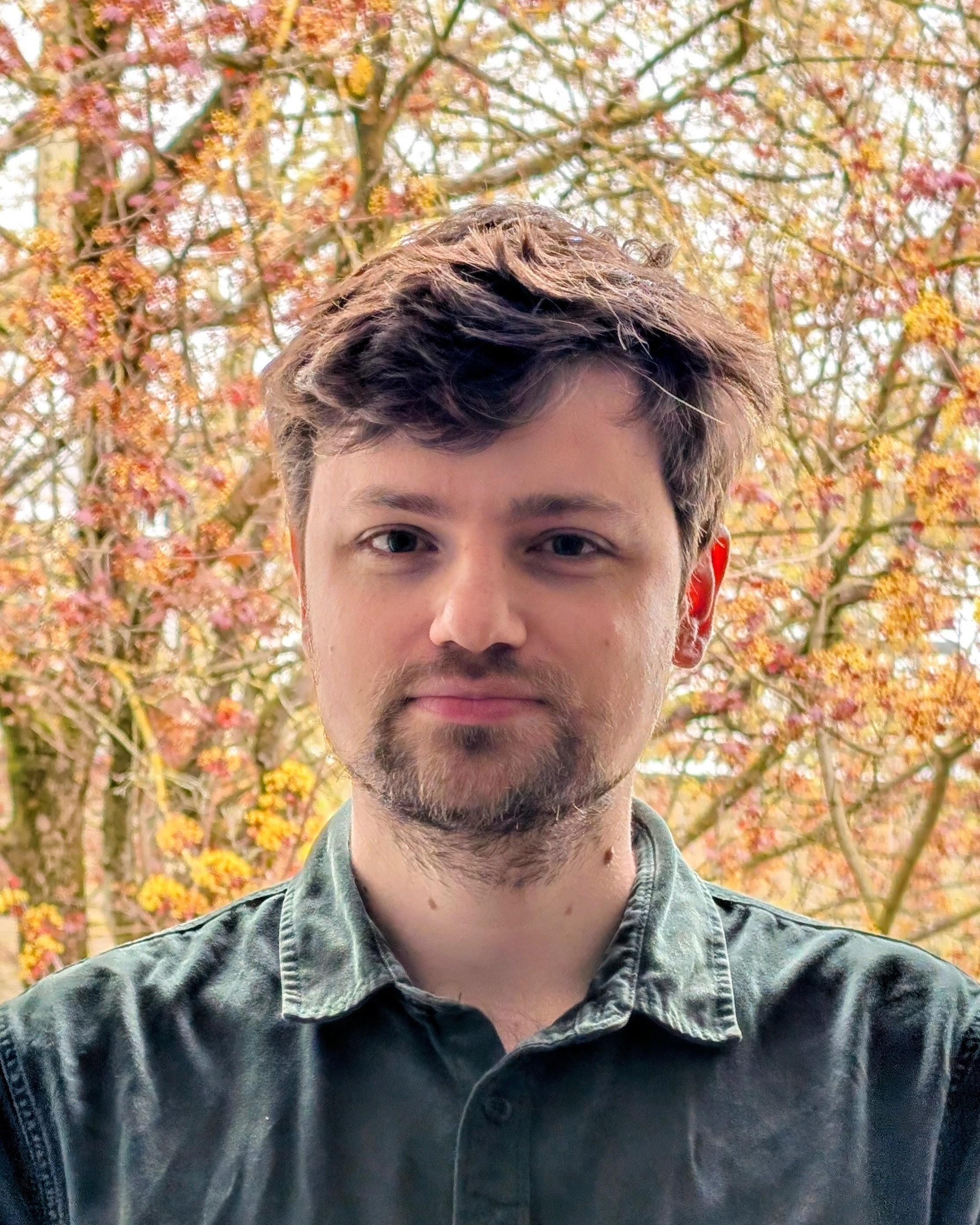}}

\begin{document}
\vspace*{4cm}
\title{New Physics in inclusive $\bar{B} \to X_c \ell \bar{\nu}$ decays}

\author{\underline{Alexandre Carvunis$^a$}, Gael Finauri$^b$, Paolo Gambino$^b$, Martin Jung$^b$, Sandro Mächler$^b$}

\address{\it $^a$TU Dortmund University, Department of Physics, \\
Otto-Hahn-Str.4, D-44221 Dortmund, Germany \\ 
\it $^b$Dipartimento di Fisica, Università di Torino \& INFN, Sezione di Torino,\\
Via Pietro Giuria 1, I-10125 Turin, Italy}

\maketitle\abstracts{
We present the first global phenomenological fit of inclusive $\bar B\to X_c\ell\bar\nu$ observables to all available experimental data allowing for generic dimension-six New Physics interactions in the Weak Effective Theory.
The fit includes both New Physics Wilson coefficients and non-perturbative QCD parameters. 
New Physics contributions are calculated including power corrections up to $\mathcal O(\Lambda_{\rm QCD}^3/m_b^3)$ and perturbative QCD corrections up to $\mathcal O(\alpha_s)$.
We find no significant preference for New Physics and obtain bounds on the size of the relevant Wilson Coefficients competitive with those coming from exclusive modes.
}

\section{Introduction}

Semileptonic $b\to c\ell \bar\nu$ transitions play a central role in precision flavour physics. As a tree-level transition, these modes have relatively large branching ratios, which makes their accurate measurement possible in $B$-factories. 
On the theory side, the heavy-to-heavy transition can be treated in the Heavy Quark Expansion (HQE) framework for both exclusive and inclusive modes. 
HQE, combined with experimental or lattice inputs helps providing theory predictions with a clear handle on uncertainties. 

Yet, several long-standing discrepancies persist in the phenomenology of $b\to c\ell \bar{\nu}$ transitions in the Standard Model, like in the observable $R(D^{(*)})$ and in the extraction of the CKM matrix element $|V_{cb}|$. 
Along with other anomalies in $B$ physics, notably in $b \to s \mu \mu$ transitions, this has triggered speculations about the presence of New Physics (NP) in the beauty sector. 

As the phenomenology of inclusive $\bar B\to X_c\ell\bar\nu$ observables reaches the $\mathcal{O}(1\%)$ accuracy in the Standard Model~\cite{Finauri:2023kte} (SM), it becomes a tantalizing probe for physics beyond the Standard Model.
Besides their absolute sensitivity to New Physics, inclusive observables probe different combinations of Wilson coefficients than their exclusive counterparts and therefore provide complementary information~\cite{Jung:2018lfu}.

Previous work on this matter has not considered fully the effect of New Physics in the extraction of the HQE parameters entering the prediction of the $\bar B\to X_c\ell\bar\nu$ observables~\cite{Blok:1993va,Manohar:1993qn,Grossman:1994ax,Feger:2010qc,Colangelo:2016ymy,Kamali:2018bdp,Fael:2022wfc}. In this proceeding we present the first simultaneous fit of NP Wilson Coefficient and HQE parameters on experimental data in a generic New Physics scenario~\cite{Carvunis:2026JHEP}.

\section{Theoretical Framework}

Including all the relevant dimension-6 operators for $b \to c \ell \bar \nu$ transitions, the Hamiltonian reads
\begin{equation}\label{eq:Hamiltonian}
\mathcal{H}_{W}= \frac{4 G_F}{\sqrt{2}} V_{cb} \sum_i C_i \mathcal{O}_i+ \rm{h.c.}\,,
\end{equation}
\bea
&\mathcal{O}_{V_L} = [\bar{c}\, \gamma^\mu P_L b] \; [\bar{\ell}\, \gamma_\mu P_L \nu_{\ell}]\,, \qquad &\mathcal{O}_{V_R} = [\bar{c}\, \gamma^\mu P_R b] \;[\bar{\ell}\, \gamma_\mu P_L \nu_{\ell}]\,, \nonumber\\
&\mathcal{O}_{S} = [\bar{c}\, b] \; [\bar{\ell}\, P_L \nu_{\ell}]\,, \qquad\qquad\quad\;\; &\mathcal{O}_{P} = [\bar{c}\, \gamma_5 b] \; [\bar{\ell}\, P_L \nu_{\ell}]\,,\nonumber\\
&\mathcal{O}_{T} = [\bar{c}\, \sigma^{\mu \nu} P_L b] \; [\bar{\ell}\, \sigma_{\mu \nu} P_L \nu_{\ell}], \quad\;\; \nonumber
\label{eq:NPopbasis}
\eea
In the SM $C_{V_L} =1 + \mathcal{O}(\alpha_{em})$ and the other Wilson Coefficients (WCs) are zero. The total differential decay rate $d\Gamma (\bar B \to X_c \ell \bar{\nu}_\ell)$ is proportional to $\sum_{X_c} \vert \left< X_c \ell \bar{\nu}_\ell \vert \mathcal{H}_W \vert \bar{B} \right> \vert^2$ where the sum runs over all once charmed hadronic state. Neglecting $\mathcal{O}(\alpha_{em})$ contributions, it can be written in the form
\begin{equation}
    \frac{d\Gamma (\bar{B} \rightarrow X_c \ell \bar{\nu})}{d E_\ell d E_{\bar{\nu}} d q^2} = \;64\pi G_F^2 |V_{cb}|^2 \frac{d^3 p_\ell}{\left(2\pi \right)^3 2E_\ell} \frac{d^3 p_{\bar{\nu}}}{\left(2\pi \right)^3 2E_{\bar{\nu}}} \sum_{\Gamma,\Gamma'} L_{\Gamma,\Gamma'} W_{\Gamma,\Gamma'}\,,
\end{equation}
where $\Gamma^{(\prime)} \in \{1, \gamma^\mu, \sigma^{\mu\nu} \}$ and 
\begin{equation}
    \label{eq:lepthadrtensDef}
L_{\Gamma,\Gamma'} = \frac{1}{4} \sum_{\rm{spins}} \left\langle \bar{\nu} \ell \left| \bar{\ell}\, \Gamma P_L \,\nu_\ell \right| 0 \right\rangle \left\langle \bar{\nu} \ell \left| \bar{\ell} \,\Gamma' P_L\, \nu_\ell \right| 0 \right\rangle^\dagger,
\end{equation}
which is calculable exactly. We also define the hadronic tensor as
\begin{equation}
\label{eq:lepthadrtensDef}
W_{\Gamma,\Gamma'} = \frac{\left(2\pi \right)^3}{2m_B} \sum_{X_\mathrm{c}} \delta^4 (p_B - p_{X_c} - q) \left\langle X_c \left | \bar{c} \,\Gamma d_\Gamma \, b \right| \bar{B} \right\rangle \left\langle X_c \left|\bar{c} \,\Gamma' d_{\Gamma^\prime}\, b \right| \bar{B} \right\rangle^\dagger\,,
\end{equation}
where we defined $q = p_\ell + p_{\bar{\nu}}$, and $d_{\Gamma^{\prime}}$ are matrices in spinor space defined as
\begin{equation}
    d_{\gamma_\mu} = C_{V_L} P_L + C_{V_R} P_R\,, \qquad d_{1} = C_S + C_P \gamma_5\,, \qquad d_{\sigma_{\mu \nu}} = C_T P_L\,.
\end{equation}
The hadronic tensor can be calculated using the optical theorem~\cite{Shifman:1985}
\begin{equation}
W_{\Gamma, \Gamma^\prime} = \frac{i}{2\pi} \textrm{disc} [T_{\Gamma, \Gamma^\prime}]\,,
\end{equation}
where we introduced the forward-scattering amplitude defined as
\begin{equation}
\label{eq:TDef}
T_{\Gamma,\Gamma'} = -\frac{i}{2m_B} \int d^4x e^{-iq\cdot x}\langle \bar{B} | \textrm{T} \{J_{\Gamma'}^\dagger(x) J_{\Gamma}(0)\} | \bar{B} \rangle \,.
\end{equation}
with the quark currents are defined as $J_\Gamma = \bar{c} \,\Gamma d_{\Gamma}\, b$. Thanks to the hierarchy $m_b\gg\Lambda_{\rm QCD}$, an operator product expansion provides a systematic expansion of the forward scattering matrix element in 
\begin{equation}
T_{\Gamma, \Gamma^\prime} = T_{\Gamma, \Gamma^\prime}^{\rm LP} + \frac{\mu_\pi^2}{m_b^2}T_{\Gamma, \Gamma^\prime}^{\mu_\pi} + \frac{\mu_G^2}{m_b^2}T_{\Gamma, \Gamma^\prime}^{\mu_G} +\frac{\rho_{LS}^3}{m_b^3}T_{\Gamma, \Gamma^\prime}^{\rho_{LS}} +\frac{\rho_{D}^3}{m_b^3}T_{\Gamma, \Gamma^\prime}^{\rho_{D}} + \mathcal{O}\left(\frac{\Lambda_{QCD}}{m_b} \right)^4 
\end{equation}
Where $\mu_\pi, \, \mu_G,\, \rho_D, \rho_{LS} $ are the so-called HQE parameters encoding non-perturbative QCD effects. 
These coefficient, which are $\mathcal{O}(\Lambda_{\rm{QCD}})$, are \textit{a priori} unknown and are fitted to the data, together with the Wilson Coefficients $C_i$.
Additionally, each order of the power expansion of the OPE receives perturbative QCD corrections.
We use the state-of-the-art calculation of the OPE of the forward-scattering matrix element in both the SM and the NP. 

The different expansion orders considered in this work are summarized in Table~\ref{tab:OPEorders}. Please check the source article~\cite{Carvunis:2026JHEP} for complete references. For New Physics contributions we provide the expressions for $W_{\Gamma,\Gamma^\prime}$ up to order $\mathcal{O}(1/m_b^3)$ in the generic New Physics case~\cite{Carvunis:2026JHEP}, as well as the $\mathcal{O}(\alpha_s)$ corrections~\cite{Carvunis:2026NP}.
\begingroup
\renewcommand{\arraystretch}{1.5}
\begin{table}[t]
\caption[]{Orders of the OPE of the forward scattering matrix element}
\label{tab:OPEorders}
\vspace{0.4cm}
\begin{center}
\begin{tabular}{|c|l|l|}
\hline
& $\Gamma$ & $\mathcal{O}(\Lambda_{QCD}^3/m_b^3) + \mathcal{O}(\alpha_s^3) + \mathcal{O}(\alpha_s/m_b^3) + \mathcal{O}(\alpha_{em})$ \\ \cline{2-3}
SM & $\left< q^2 \right>$ & $\mathcal{O}(\Lambda_{QCD}^3/m_b^3) + \mathcal{O}(\alpha_s^2) + \mathcal{O}(\alpha_s/m_b^3) $ \\ \cline{2-3}
& $R^*, \, \left< E_\ell \right>, \, \langle m_{X_c}^2 \rangle$ & $\mathcal{O}(\Lambda_{QCD}^3/m_b^3) + \mathcal{O}(\alpha_s^2) + \mathcal{O}(\alpha_s/m_b^2) $ \\ \hline
\multicolumn{2}{|c|}{NP} & $ \mathcal{O}(\Lambda_{QCD}^3/m_b^3) + \mathcal{O}(\alpha_s)$ \\ \hline
\end{tabular}
\end{center}
\end{table}
\endgroup

\section{Observables and fit procedure}

The result provided by the calculation introduced above provides an unphysical fully differential decay rate. This follows directly from the observation that, at $\mathcal{O}(\alpha_s^0)$, power corrections of order $\mathcal{O}(\Lambda_{\mathrm{QCD}}^n/m_b^n)$ are proportional to derivatives of Dirac delta distributions, $\delta^{(n-1)}[(m_b v - q)^2 - m_c^2]$. In fact, the OPE is only valid when the differential decay rate is integrated over a wide enough range of the phase space. For this reason, the observables of choice for $\bar B \to X_c \ell \bar \nu$ are the normalized moments of the decay rate defined as 
\begin{equation}
\mathcal{M}_n = \left< X^n \right> = \frac{\int dX \,\frac{d\Gamma}{d X} X^n }{\int dX \,\frac{d\Gamma}{d X}}
\qquad
X=E_\ell,\; m_{X_c}^2,\; q^2,
\end{equation}
measured with lepton-energy or $q^2$ cuts. In practice the central moments are used to reduce correlations between moments: $\equiv \left< (X - \left< X \right>)^n \right>$ for $n \geq 2$. We provide results for $n=1,2,3$ as moments above $n \geq 4$ do not provide meaningful constraints. Our results include for the first time analytical formulas for the $\mathcal{O}(\alpha_s)$ corrections to kinematic moments in the presence of New Physics~\cite{Carvunis:2026NP}. The other integrated observable used for the fit is
\begin{equation}
    R^*(E_{\ell \rm{cut}}) \equiv \frac{\textrm{BR}_{E_{\ell \rm{cut}}}(\bar B \to X_c \ell \bar \nu)}{\textrm{BR}(\bar B \to X_c \ell \bar \nu)}.
\end{equation}

We use all available experimental data measured in $B-$factories, namely: Babar, CLEO, DELPHI, CDF, Belle and Belle II. In the SM, the fit procedure goes in two steps: 
\begin{enumerate}
    \item Fit the ratio observables - independent of $V_{cb}$ - to experimental data. The fitted parameters are:
    \bea
         m_b, \, m_c, \, \mu_\pi^2,\mu_G^2,\rho_{LS}^3,\rho_D^3, \textrm{BR}(\bar B \to X_c \ell \bar \nu) \nonumber 
    \eea
    \item Extract $V_{cb}$ using the total decay rate following an expression of the form
    \begin{equation}
        \Gamma_{B \to X_c\ell \nu} = \vert V_{cb} \vert^2 f(m_b,m_c,\mu_\pi^2,\mu_G^2,\rho_{LS}^3,\rho_D^3) = \frac{\textrm{BR}(\bar B \to X_c \ell \bar \nu)}{\tau_B}
    \end{equation}
\end{enumerate}
We provide an updated extraction of $\vert V_{cb} \vert$ in the SM, including for the first time $\mathcal{O}(\alpha_s^2)$ corrections in the $q^2$-moments~\cite{Fael:2024gyw} and the full $\mathcal{O}(\alpha_{em})$~\cite{Bigi:2023cbv} correction in the total decay rate. We find~\cite{Carvunis:2026JHEP}
\begin{equation}
    \vert V_{cb} \vert = 41.64(47) \times 10^{-3}
\end{equation}
When including New Physics contributions, a parametric degeneracy arises between $V_{cb}$ and the Wilson coefficients $C_i$, as seen in (\ref{eq:Hamiltonian}). To get rid of this issue we remove one degree of freedom by defining
\begin{equation}
\tilde V_{cb} \equiv V_{cb} C_{V_L}.
\end{equation}
In the limit $m_\ell = 0$, we can parameterize the full New Physics effects with $7$ parameters. We work with a polar representation of the normalized WCs
\begin{equation}
\frac{C_{X}}{C_{V_L}} \equiv a_{X} e^{i \delta_{X}}, \quad X = V_R, T, S, P
\end{equation}
and use the following fit parameters: $a_{V_R}, a_T, a_P, a_S, cos(\delta_{V_R}), cos(\delta_S - \delta_T), cos(\delta_P - \delta_T)$ in addition to the SM parameters already introduced.
\section{Fit results}

\begin{figure}
    \centering
    \includegraphics[width=\linewidth]{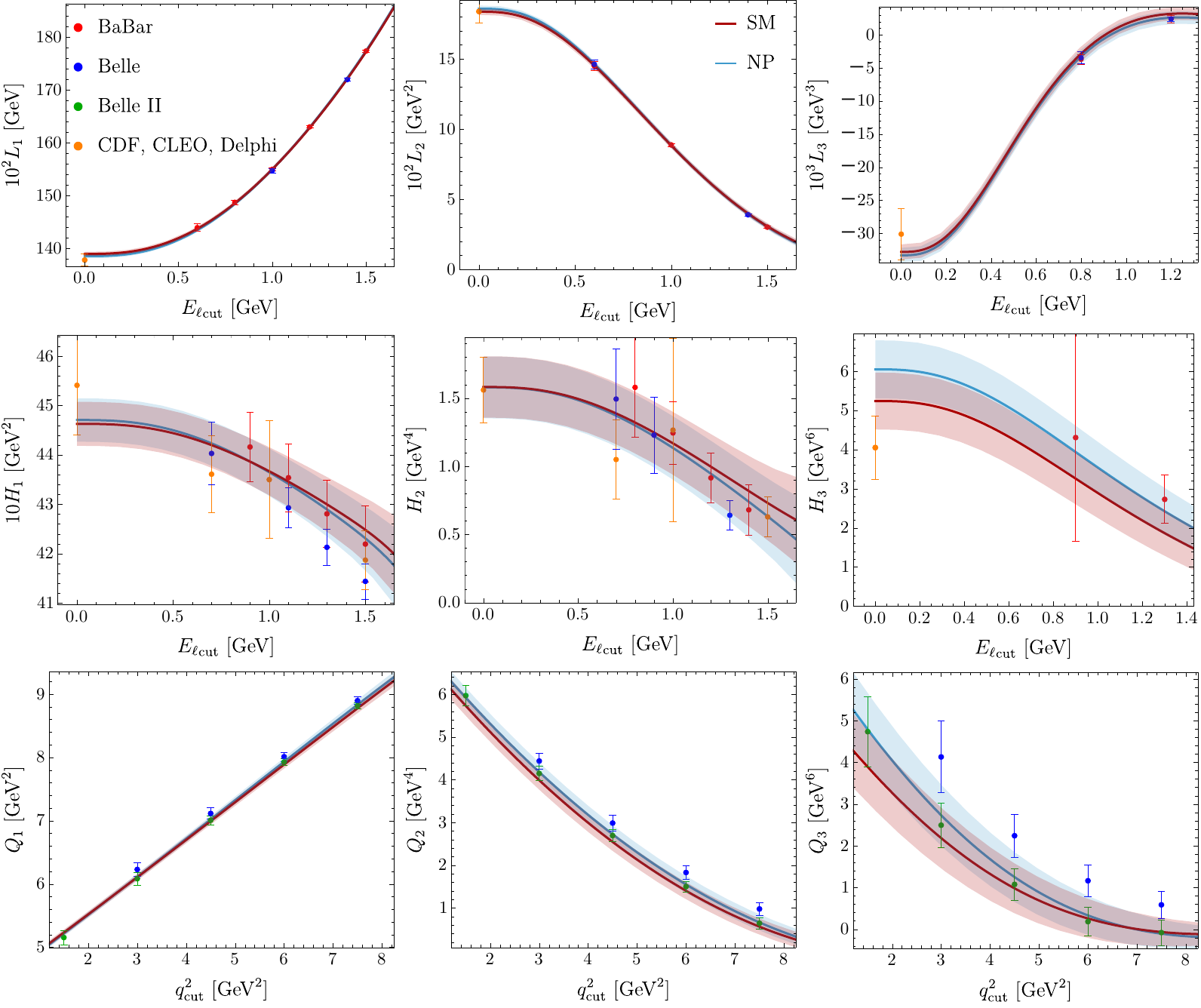}
    \caption{Observables included in the fit, with the theoretical predictions obtained from the SM (red) and full NP (blue) fit results. The shaded bands denote the theoretical uncertainties.}
    \label{fig:observablefit}
\end{figure}

We obtain a good fit to data in both the SM and the NP case with $\chi^2_{\rm SM}/{\rm d.o.f.}=42.6/74$ and $\chi^2_{\rm NP}/{\rm d.o.f.}=36.1/67$. 
This is illustrated in Fig.~\ref{fig:observablefit} which shows the SM and NP theoretical predictions fitted to data. 

The NP hypothesis is compatible with the SM at the $0.7 \sigma$ level, which was expected given the goodness of the SM fit. In Fig.~\ref{fig:param_fullNP} we show the 2D profiled $\chi^2$ allowing for all New Physics parameters. One can see that all the New Physics fit parameters are compatible with zero. However the fit provides an important piece of information, it provides upper bounds on the magnitude of the WCs competitive with bounds from exclusive decays.
\begin{figure}
    \centering
    \includegraphics[width=\linewidth]{img/profile2D_fullNP.pdf}
    \caption{2D contour plots for pairs of fit parameters in the global fit with $\Delta \chi^2 \leq 2.30$ (dark blue) and $\Delta\chi^2 \leq5.99$ (light blue)
    (plots for the phases are not shown for simplicity). The red dot indicates the best fit point.}
    \label{fig:param_fullNP}
\end{figure}

A quirk of our result is the profiled value of $|\tilde V_{cb}|$, we find
\begin{equation}
|\tilde V_{cb}|=
(35.3^{+4.4}_{-3.6})\times10^{-3}.
\end{equation}
which has a very low central value, but is still compatible with the exclusive and inclusive determinations thanks to the large uncertainties. 
The relatively large uncertainty originates from approximate flat directions in the kinematic moments, which are absent from the total decay rate.
The latter flat direction can be easily removed by including exclusive modes to the fit. 
One could also address this issue within the framework of inclusive decays by measuring observables optimized for this purpose in Belle II. 

\section{Outlook}

In this work we present the first global fit of $\bar{B} \to X_c \ell \bar{\nu}$ observables with model-independent New Physics and show that they provide new and competitive bounds on New Physics in $b \to c \ell \bar{\nu}$ transitions~\cite{Carvunis:2026JHEP}.

In addition, we provide the first fully analytical expressions of the $\mathcal{O}(\alpha_s)$ corrections to the first three kinematic moments in $E_\ell, m_{X_c}^2$ and $q^2$~\cite{Carvunis:2026NP}. This simplifies greatly the utilization of our results in other works of phenomenology and provides a performance advantage in numerical computations.

The next logical step of this work is to combine bounds from exclusive and inclusive decay modes in $b \to c \ell \bar{\nu}$. 

\section*{Acknowledgments}

The work of AC is supported by the \textit{Bundesministerium f\"ur Forschung, Technologie und Raumfahrt } -- BMFTR.
The work of GF, PG and MJ is supported by the Italian Ministry of University and Research (MUR) and the European Union (EU) – Next Generation EU, Mission 4, Component 1, PRIN 2022, grant 2022N4W8WR, CUP D53D23002830006.

\section*{References}



\end{document}